\documentclass[english,amsmath,amssymb,prd,nofootinbib]{revtex4-1}
\usepackage[T1]{fontenc}
\usepackage[latin9]{inputenc}
\usepackage[letterpaper]{geometry}
\usepackage{color}
\usepackage{babel}
\usepackage{amssymb}
\usepackage{graphicx}
\usepackage{stmaryrd}
\usepackage{mathtools}
\usepackage{enumerate}
\usepackage{xcolor}

\newcommand{\be}{\begin{equation}}
\newcommand{\ee}{\end{equation}}
\newcommand{\bea}{\begin{eqnarray}}
\newcommand{\eea}{\end{eqnarray}}
\newcommand{\xdot}{\dot{x}}

\begin{document}

\title{Realization of DSR-relativistic symmetries in Finsler geometries}

\newcommand{\addressSISSA}{SISSA, Via Bonomea 265, Trieste, 34136, Italy and INFN, Sezione di Trieste}
\newcommand{\addressRoma}{Dipartimento di Fisica, Universit\`a ``La Sapienza''
and Sez. Roma1 INFN, P.le A. Moro 2, 00185 Roma, Italy}
\newcommand{\addressPI}{Dipartimento di Matematica, Universit\`{a}  "La Sapienza", P. le A. Moro 2, 00185 Roma, Italy and Perimeter Institute for Theoretical Physics, 31 Caroline Street North, Waterloo, ON, N2L2Y5 Canada}

\author{Giovanni Amelino-Camelia}
\affiliation{\addressRoma}
\author{Leonardo Barcaroli}
\affiliation{\addressRoma}
\author{Giulia Gubitosi}
\affiliation{\addressRoma}
\author{Stefano Liberati}
\affiliation{\addressSISSA}
\author{Niccol\'o Loret}
\affiliation{\addressPI}

\begin{abstract}
Finsler geometry is a well known generalization of Riemannian geometry which allows to account for a possibly non trivial structure of the space of configurations of relativistic particles.
We here establish a link between Finsler geometry and the sort of models with curved momentum space and DSR-relativistic symmetries which have been recently of interest in the quantum-gravity literature. We use as case study the much-studied scenario which is inspired by the  $\kappa$-Poincar\'e quantum group, and show that the relevant deformation of  relativistic symmetries can be implemented within a Finsler geometry.
\end{abstract}

\maketitle
\nopagebreak
\tableofcontents

\section{Introduction}

Our main objective here is to uncover an apparently deep connection between two much-studied theoretical-physics frameworks, the one of Finsler geometry and the one of DSR-relativistic symmetries.

Theories with two non-trivial relativistic invariants, also known as ``DSR-relativistic theories" \cite{GACijmpd2002, smolinDSR2003, kowalskiDSR2003, gacSYMMETRYreview},  have attracted much interest in the recent quantum-gravity literature.
The presence of the second relativistic invariant  has been shown to be related to curvature in momentum space,  with the scale of curvature of momentum space (tentatively expected to be of the order of the Planck scale) playing a role completely analogous to the one already attributed to the speed-of-light scale within special relativity.
Examples of DSR-relativistic theories  can be inspired by the study of quantum groups, in which case momentum
space turns out to be a group manifold. They are believed to be an indirect
manifestation of spacetime quantization, strongly suggesting  that spacetime may not be described by a Riemannian geometry at ultra-short scales.

An even richer stream of research characterizes the study of Finsler geometry, which could be viewed as an approach suitable for  abandoning Riemannian geometry as the arena for the relativistic dynamics of particles, essentially allowing for a velocity dependent geometry to describe spacetime structure.

There is already a well-established common point between the DSR-relativistic theories framework and Finsler geometry, and this is the possibility of allowing for modified dispersion relations. It was already shown in \cite{MDRFinsler} that Finsler metrics can be used to describe the geometry on which a particle with modified dispersion relation lives, but it was not investigated whether Finsler geometries have room to accommodate also a description of the (modified) relativistic symmetries.

We tackle this challenge here by focusing on the illustrative
example of the curved momentum space that was inspired \cite{GiuliaFlaviokappaLocality, Trevisan}
by the so-called $\kappa$-Poincar\'e quantum-group deformation of the Poincar\'e group \cite{Lukierski:1992dt, Lukierski:1993wxa, Lukierski:1991pn}.
$\kappa$-Poincar\'e is a widely studied candidate
to describe departures from Special Relativity that could arise
in a "semiclassical" regime of Quantum Gravity (whose scale is set by the value of $\kappa$),
where the gravitational degrees of freedom are integrated out, leaving an effective field theory for matter. Indeed this was shown to be the case at least for $2+1$-dimensional quantum gravity \cite{Freidel:2005me}.

An important player in our analysis is a known  prescription for deriving the Killing vectors associated to a given Finsler geometry. Until now it was not clear whether these Killing vectors are actually associated to  symmetries that leave invariant a given dispersion relation. For the $\kappa$-Poincar\'e-inspired momentum space, which we use as illustrative example of our thesis, the relationship between (modified) dispersion and relativistic transformations has been already studied in detail and this placed us in a strong position for investigating a possible description in terms of Finsler geometry.

Without loosing any of the conceptual challenges that are here of interest we work in $1+1$ dimensions and at the first order in the deformation parameter $\ell$ ($\ell\sim 1/\kappa$ is a length scale, related to the scale of curvature of momentum space), so that formulas are less bulky and indeed the conceptual issues come more to the forefront. Our greek indices have values $\{0,1\}$, and  we set $c$ and $\hbar$ to one.

\section{Description of particles with $\kappa$-Poincar\'e symmetries}
\label{sec:kappaPoincare}

The $\kappa$-Poincar\'e group \cite{Lukierski:1992dt, Lukierski:1993wxa, Lukierski:1991pn} is a deformation of the Poincar\'e group that accommodates a second invariant scale (an energy scale) besides the speed of light, without violating the relativity principle.  The energy scale $\kappa$ governs the departures from the standard special relativistic symmetries. We will indicate it as  $1/\ell$, where $\ell$ is a parameter with dimensions of a length,  expected to be of the order of the Planck length.

When the so-called bicrossproduct basis \cite{MajidBicross} is chosen, the $\kappa$-Poincar\'e generators associated to spacetime translations ($P_{0}, P_{1}$) and boost ($\mathcal{N}$) satisfy the following Lie brackets \footnote{Because of the classical nature of the physical framework that we are going to study (there are no \emph{pure} quantum effects, i.e. $\hbar \sim 0$) we use as Lie brackets the Poisson ones.}:
\begin{eqnarray}
&& \{P_0,P_1\} = 0 \nonumber \\
&& \{\mathcal{N},P_0\} = P_1\label{eq:kappaAlgebra}  \\
&& \{\mathcal{N},P_1\} = P_{0}-\ell P_{0}^{2} - \frac{\ell}{2}P_1^2.\nonumber
\end{eqnarray}
The Casimir of this algebra reads:
\be
\mathcal{C}_\ell = P_0^2 -P_1^2 -\ell P_0 P_1^2. \label{eq:Casimirk}
\ee

One can obtain a finite transformation from the set of infinitesimal transformations described by the generators by means of the exponential
map \cite{lateshift}:
\begin{equation}
\mathcal{F} \rhd f(x,p) = \sum_{n=0}^{\infty} \frac{1}{n!} \underbrace{\{d^\mu p_\mu + a \mathcal{N},\{\dots,\{d^\mu p_\mu + a \mathcal{N}}_{\text{n times}},f(x,p)\}\dots\}\},
\end{equation}
where $a$ and $d^\mu$ are respectively the (finite) boost and translations parameters.

Upon choosing the trivial (Darboux) symplectic structure for the phase  space\footnote{At the quantum level the $\kappa$-Poincar\'e group is  related to the $\kappa$-Minkowski noncommutative spacetime \cite{MajidBicross,  Lukierski:1993wx}, but for classical phase-space constructions one can consider standard Minkowski spacetime coordinates, related to the noncommutative coordinates by a momentum-dependent redefinition \cite{Rosati:2012fb}.}
\begin{eqnarray}
&& \{x^\mu,x^\nu\} = 0 \nonumber\\
&& \{x^\mu,p_\nu\} = \delta^\mu_\nu\label{darboux}\\
&& \{p_\mu,p_\nu\} = 0 \nonumber
\end{eqnarray}
one finds that the symmetry generators have the following representation:
\bea
&&P_{0}=p_{0}, \quad P_{1}=p_{1},\nonumber\\
&&\mathcal N=x^{0} p_{1}+x^{1} \left(p_{0}-\ell p_{0}^{2}-\frac{\ell}{2}p_{1}^{2} \right),
\eea
This representation ensures that the generators have vanishing Poisson brackets with the Casimir and is also compatible with the following Poisson brackets between boost and coordinates:
\begin{equation}
\{{\cal N},x^0\}=-x^1(1-2\ell p_0)\;,\;\;\;\{{\cal N},x^1\}=-x^0+\ell p_1 x^1\,.\label{eq:BoostCoordParentheses}
\end{equation}

The dispersion relation of a relativistic particle can be then deduced from the representation of the Casimir as:
\begin{equation}
m^{2}=\mathcal{C}_\ell(p) \equiv p_0^2 -p_1^2 -\ell p_0 p_1^2.\label{eq:dSdispersionRelation}
\end{equation}

The prescription for deriving a particle worldline relies on the Hamiltonian formalism using the particle Casimir $\mathcal C_{\ell}$ as Hamiltonian \cite{lateshift, AmelinoCamelia:2010qv, kbob, Rosati:2012fb}:
\begin{equation}
\dot{p}_\mu=\{p_\mu,\mathcal{C}_\ell\} 
\label{alg_eom_p}
\end{equation}
\begin{equation}
\dot{x}^\mu = \{x^\mu,\mathcal{C}_\ell\} 
\label{alg_eom_x}
\end{equation}
where the dot represents derivation with respect to the affine parameter on the worldline.
From these equations one deduces energy/momentum conservation along the worldline
\be
\dot p_\mu=0,
\ee
and a differential equation for the coordinates along the worldline:
\begin{eqnarray}
&& \dot{x}^0 = 2 p_{0}-\ell p_{1}^{2} \\
&& \dot{x}^1 = -2  p_1(1+\ell p_{0})\label{eq:dotx1}
\end{eqnarray}

After solving the differential equations and exploiting the dispersion relation one can eliminate the affine parameter and find:
\be
x^{1}-\bar x^{1}=-\frac{\sqrt{p_{0}^{2}-m^{2}}}{p_{0}}\left(1+\ell\frac{ (2 p_{0}^{2}-m^{2})}{2 p_{0}}\right)(x^{0}-\bar x^{0}) \label{eq:dSworldlineMassive}
\ee
where $\{\bar x^{0},\bar x^{1}\}$ are the starting coordinates of the worldline.

A few remarks are in order before we close this  section.
A reader familiar with the quantum groups theory will have noticed that we are here completely neglecting the co-algebraic structure of $\kappa$-Poincar\'e. This is because this work focuses on a single-particle system, where the co-products of the symmetry generators are not relevant.

As mentioned in the introduction, a geometrical interpretation of the $\kappa$-Poincar\'e group was  proposed recently \cite{GiuliaFlaviokappaLocality, Trevisan}, based on the observation \cite{kowalskiDSR2003, KowalskiGlikman:2003we, KowalskiGlikman:2004tz} that the translation generators of $\kappa$-Poincar\'e live on a curved momentum manifold. Indeed the symmetry generators of $\kappa$-Poincar\'e leave invariant the momentum-space line element:
\be
ds^2=\zeta^{\mu\nu}(p)dp_\mu dp_\nu,
\ee
where the momentum space metric $\zeta^{\mu\nu}(p)$ reads:
\begin{equation}
\zeta^{\mu\nu}(p)=
\begin{pmatrix}
1 & 0 \\
0 & -(1+2 \ell p_0)
\end{pmatrix}.\label{eq:momSpacemetric}
\end{equation}
This is clearly the metric of a de Sitter manifold written in ``flat slicing coordinates'' and at the first order in $\ell$.
We are not giving here the details of the momentum space geometrical structure induced  by the $\kappa$-Poincar\'e group and of its physical interpretation, as in this manuscript  we are concerned with analyzing the modifications on the spacetime structure that are required in order to accommodate modified relativistic symmetries such as the ones described by $\kappa$-Poincar\'e.

\section{Finsler geometry of a particle with modified dispersion relation}

In \cite{MDRFinsler} a prescription for deriving the Finsler geometry of a particle with modified dispersion relation and living on a flat spacetime was provided. We report here the main steps of the construction. For details about the construction and about Finsler geometry in general we refer the reader to \cite{MDRFinsler} and references therein.

The starting point is the action of a particle with modified dispersion relation of the form $m^{2}=\mathcal M(p)$:
\be
I=\int \left(\dot x^{\mu}p_{\mu}-\lambda\left(\mathcal M(p)-m^{2}\right)\right)d\tau. \label{eq:LagrangianFinsler}
\ee
Here $\lambda$ is a Lagrange multiplier that enforces the mass-shell condition and the over-dot stands for derivation with respect to the affine parameter $\tau$.

By using  Hamilton's equation
\be
\dot x^{\mu}=\lambda \frac{\partial \mathcal M}{\partial p_{\mu}}, \label{eq:EOMfromLagrangian}
\ee
one can find the relation between momenta $p_{\mu}$ and velocities $\dot x^{\mu}$ and write the action as a function of velocities and the Lagrange multiplier:
\be
I=\int \mathcal L(\dot x,\lambda)d\tau.
\ee
Then by varying the action with respect to $\lambda$ one can substitute $\lambda\rightarrow \lambda(\dot x)$ and obtain the final form of the action as \footnote{We will show in section \ref{sec:LineElement} that the Lagrangian appearing in this action is invariant under boosts only up to total derivatives. This is not worrisome, since the action is invariant and gives covariant equations of motions. Despite this, one might want to write the action in terms of an invariant Lagrangian, and in section \ref{sec:LineElement} we will  show that this is indeed possible, redefining the Lagrangian so that it is  invariant and still gives the same (covariant) worldlines as this one.}:
\be
I=\int \mathcal L(\dot x,\lambda(\dot x))d\tau.
\ee
The Lagrangian written as a function of velocities only  can be now identified with a Finsler norm,
\be
\mathcal L(\dot x,\lambda(\dot x))\equiv m F(\dot x), \label{eq:NormFromLagrangian}
\ee
 since it satisfies the required properties \footnote{Note that in general a Finsler norm depends on both coordinates and velocities. However, since in this case we are considering deformations of a special-relativistic particle, which lives on a  flat spacetime, the Finsler norm is only velocity-dependent.}
\be
\left\{\begin{array}{l}
   F(\dot{x})\neq 0\;\;\; \text{if}\;\;\; \dot{x}\neq 0\\
   \;\\
   F(\epsilon\dot{x})=|\epsilon|F(\dot{x}). \label{eq:FinslerConditions}
\end{array}\right .
\ee

A  Finsler norm can in general be associated to a metric
\be
g_{\mu\nu}(x,\dot x)= \frac{1}{2}\frac{\partial^{2} F^2(x,\dot x)}{\partial \dot x^{\mu}\partial \dot x^{\nu}}.\label{eq:FinslerMetricDefinition}
\ee
The metric will only be a function of velocities if the Finsler norm does not depend on $x^{\mu}$ (which is the case of interest in this work).

Using the Euler theorem for homogeneous functions applied to $F^2$:
\be
\dot x^{\mu} \frac{\partial F^{2}}{\partial \dot x^{\mu}}=2 F^{2}, \label{eq:EulerTheoremF2}
\ee
one can show that the metric just defined satisfies:
\be
\dot x^{\alpha} \frac{\partial g_{\mu\nu}}{\partial \dot x^{\alpha}}=\dot x^{\mu} \frac{\partial g_{\mu\nu}}{\partial \dot x^{\alpha}}=\dot x^{\nu} \frac{\partial g_{\mu\nu}}{\partial \dot x^{\alpha}}=0,\label{eq:propertiesofgF}
\ee
and
\be
F=\sqrt{g_{\mu\nu}\dot x^{\mu}\dot x^{\nu}}.\label{eq:gR}
\ee

Once the Finsler norm is given, the particle action can be written in the form
\be
I=m \int F(\dot x)  d\tau =m \int \sqrt{g_{\mu\nu}(\xdot) \xdot^\mu\xdot^\nu},
\ee
which looks like a straightforward generalization of the action of a special-relativistic particle.

The construction of the metric $g_{\mu\nu}(\dot x)$ allows to relate momenta to velocities in a simple way:
\begin{equation}
p_\mu = m \frac{\partial F}{\partial \xdot^\mu}= m \frac{g_{\mu\nu}\xdot^\nu }{F}.
\label{eq:Fmomenta}
\end{equation}
Notice that using this definition of momenta one recovers the on-shell condition as:
\be
m^2= g^{\mu\nu}(\xdot(p)) p_\mu p_\nu \label{eq:MetricandDispersionRelation}
\ee
where $g^{\mu\nu}(\xdot(p))$ is the inverse of the metric $g_{\mu\nu}(\dot x)$.

\section{Finsler geometry of a particle with $\kappa$-Poincar\'e-inspired dispersion relation}
\label{sec:FinslerBiscross}
\subsection{Deriving the Finsler norm}
In this section we will apply the procedure just described to the case of a particle whose dispersion relation is compatible with the $\kappa$-Poicar\'e Casimir and given in (\ref{eq:dSdispersionRelation}). This means that we specialize the results  of the previous section to the case $\mathcal M(p)=\mathcal C_\ell(p)\equiv p_0^2-p_1^2-\ell p_0 p_1^2$.

The Lagrangian (\ref{eq:LagrangianFinsler}) takes the form:
\be
I=\int \left(\dot x^{\mu}p_{\mu}-\lambda\left( p_{0}^{2}-p_{1}^{2}-\ell p_{0} p_{1}^{2}-m^{2}\right)\right)d\tau.\label{eq:kappaAction}
\ee
The associated equations of motion are derived as in (\ref{eq:EOMfromLagrangian}):
\bea
\dot x^{0}&=&\lambda \left(2 p_{0}-\ell p_{1}^{2}\right),\label{eq:dotx0ofp}\\
\dot x^{1}&=&\lambda \left(-2 p_{1}-2 \ell p_{0} p_{1} \right).\label{eq:dotx1ofp}
\eea
Note that these equations are  the same (\emph{i.e.} they produce the same worldlines) as the ones we found in section \ref{sec:kappaPoincare}, if we assume that the Lagrange multiplier $\lambda$ is independent from the phase space $\{x^{\mu},p_{\mu}\}$ and that $\dot \lambda=0$ (in this case we can reabsorb it in the affine parameter).

Inverting the equations of motion (\ref{eq:dotx0ofp})-(\ref{eq:dotx1ofp}), we find the expression of the momenta $p_{\mu}$ as a function of velocities $\dot{x}^{\mu}$ and of the Lagrange multiplier $\lambda$:
\begin{eqnarray}
&& p_0=\frac{\dot{x}^0}{2\lambda}+\frac{\ell}{2}\frac{(\dot{x}^1)^2}{4\lambda^2},\nonumber\\
&& p_1 =\frac{\dot{x}^1}{-2\lambda}+\ell \frac{\dot{x}^0\dot{x}^1}{4\lambda^2}.
\label{pLambda}
\end{eqnarray}
(we remind to the reader that we are working at leading order in $\ell$).
Plugging these relations into the action (\ref{eq:kappaAction}) we get the Lagrangian
\begin{equation}
\mathcal{L} =\frac{(\dot{x}^0)^2-(\dot{x}^1)^2}{4\lambda}+\lambda m^2+\ell\frac{\dot{x}^0(\dot{x}^1)^2}{8\lambda^2},\label{lambdaLagrangian}
\end{equation}
which is then minimized with respect to $\lambda$ to get:
\begin{equation}
\lambda(\dot{x})=\frac{1}{2}\frac{\sqrt{(\dot{x}^0)^2-(\dot{x}^1)^2}}{m}+\frac{\ell}{2}\frac{\dot{x}^0(\dot{x}^1)^2}{(\dot{x}^0)^2-(\dot{x}^1)^2}. \label{lambda}
\end{equation}
This allows to find a closed expression for the relations (\ref{pLambda}) between momenta and velocities \footnote{ Note that through (\ref{eq:p0_complete}) and (\ref{eq:p1_complete}) we automatically recover the Casimir, using only algebraic relations. Indeed, from (\ref{eq:p0_complete}) and (\ref{eq:p1_complete}) we get:
\begin{eqnarray}
\frac{m\dot{x}^0}{\sqrt{(\dot{x}^0)^2-(\dot{x}^1)^2}}&=&p_0+\frac{\ell}{2}\frac{p_1^2}{m^2}\left( p_0^2+p_1^2\right),\label{eq:x0dotVSp}\\
\;\nonumber\\
\frac{m\dot{x}^1}{\sqrt{(\dot{x}^0)^2-(\dot{x}^1)^2}}&=& - p_1\left(1+\frac{\ell}{m^2} p_0^3\right).\label{eq:x1dotVSp}
\end{eqnarray}
Then, subtracting the square of these two equations:
\begin{equation}
m^2=\left(\frac{m\dot{x}^0}{\sqrt{(\dot{x}^0)^2-(\dot{x}^1)^2}}\right)^2-\left(\frac{m\dot{x}^1}{\sqrt{(\dot{x}^0)^2-(\dot{x}^1)^2}}\right)^{2}=p_0^2-p_1^2-\ell p_0 p_1^2
\end{equation}}:
\begin{eqnarray}
p_0 &=& \frac{m\dot{x}^0}{\sqrt{(\dot{x}^0)^2-(\dot{x}^1)^2}} - m^2\frac{\ell}{2}\frac{(\dot{x}^1)^2((\dot{x}^0)^2+(\dot{x}^1)^2)}{((\dot{x}^0)^2-(\dot{x}^1)^2)^2},\label{eq:p0_complete}\\
\;\nonumber\\
p_1 &=& -\frac{m\dot{x}^1}{\sqrt{(\dot{x}^0)^2-(\dot{x}^1)^2}} + m^2\ell\frac{\dot{x}^1(\dot{x}^0)^3}{((\dot{x}^0)^2-(\dot{x}^1)^2)^2},\label{eq:p1_complete}
\end{eqnarray}
and write the Lagrangian as a function of velocities only:
\begin{equation}
\mathcal{L}=m\left(\sqrt{(\dot{x}^0)^2-(\dot{x}^1)^2}+\frac{\ell}{2} m\frac{\dot{x}^0(\dot{x}^1)^2}{(\dot{x}^0)^2-(\dot{x}^1)^2}\right).
\label{eq:dSFinslerLagrangian}
\end{equation}
Then the Finsler norm associated to a particle with $\kappa$-Poincar\'e-compatible dispersion relation turns out to be, through (\ref{eq:NormFromLagrangian}):
\be
F(\dot x)=\left(\sqrt{(\dot{x}^0)^2-(\dot{x}^1)^2}+\frac{\ell}{2} m\frac{\dot{x}^0(\dot{x}^1)^2}{(\dot{x}^0)^2-(\dot{x}^1)^2}\right)
\label{eq:dSFinslerNorm}
\ee

It is straightforward to verify that the function $F(\dot{x})$ just defined is a legitimate Finsler norm, since it satisfies the conditions (\ref{eq:FinslerConditions}). Notice that, as expected since we are considering the deformation of a special-relativistic particle, in this case the norm only depends on velocities and not on coordinates, meaning that the spacetime geometry is flat (see next subsection).

\subsection{Finsler metric}
\label{sec:FinslerMetric}

A Finsler metric $g_{\mu\nu}(x,\dot{x})$ can be derived from the norm (\ref{eq:dSFinslerNorm}) via the relation (\ref{eq:FinslerMetricDefinition}):
\begin{equation}
\textstyle{g_{\mu\nu}(x,\dot x)}={\large\begin{pmatrix}
1+\frac{3}{2}\ell m\frac{\dot x^{0}  (\dot x^{1})^{4}}{\left((\dot x^{0})^{2}-(\dot x^{1})^{2}\right)^{5/2}}&
\ell\frac{m}{2} \frac{-4 (\dot x^{0})^{2}(\dot x^{1})^{3}+(\dot x^{1})^{5} }{\left((\dot x^{0})^{2}-(\dot x^{1})^{2}\right)^{5/2}}\\
&\\
\ell\frac{m}{2} \frac{-4 (\dot x^{0})^{2}(\dot x^{1})^{3}+(\dot x^{1})^{5} }{\left((\dot x^{0})^{2}-(\dot x^{1})^{2}\right)^{5/2}}&
-1+\frac{1}{2}\ell m (\dot x^{0})^{3}\frac{2 (\dot x^{0})^{2}+  (\dot x^{1})^{2}}{\left((\dot x^{0})^{2}-(\dot x^{1})^{2}\right)^{5/2}}
\end{pmatrix}}
\label{fmetric}
\end{equation}

Since the Finsler norm we are considering does not  depend on coordinates $x^{\mu}$, also the metric is coordinate-independent. When the deformation parameter $\ell$ vanishes the metric reduces to the one of special relativity, and in this sense we say that the metric is flat. In the following we will see that this metric also has vanishing (generalized) Christoffel symbols.

It is easy to verify that the metric $g_{\mu\nu}$ satisfies the properties in (\ref{eq:propertiesofgF}), and as a consequence:
\begin{equation}
\dot{x}^\alpha \frac{\partial g_{\mu\nu}}{\partial \dot{x}^\alpha}\dot{x}^\mu\dot{x}^\nu=0.\label{improper_hom}
\end{equation}

We can also write the metric in terms of momenta by use of (\ref{eq:x0dotVSp}) and (\ref{eq:x1dotVSp}):
\begin{equation}
\textstyle{g_{\mu\nu}(x,p)}={\large\begin{pmatrix}
1+\frac{3}{2}\ell  \frac{p_{0}\, p_{1}^{4}}{m^{4}}&
\frac{\ell}{2} \frac{4 p_{0}^{2}\,p_{1}^{3}-p_{1}^{5} }{m^{4}}\\
&\\
\frac{\ell}{2} \,\frac{4 p_{0}^{2}\,p_{1}^{3}-p_{1}^{5} }{m^{4}}&
-1+\frac{\ell}{2} p_{0}^{3}\,\frac{2 p_{0}^{2}+p_{1}^{2}}{m^{4}}
\end{pmatrix}}\label{fmetricp}
\end{equation}

When we use this expression of the metric in terms of momenta we find a simple relation with the particle dispersion relation (note that we use here the inverse of the metric):
\bea
g^{\mu\nu}(x,p)p_{\mu}p_{\nu}&=& p_{0}^{2}-p_{1}^{2}-\ell p_{0} p_{1}^{2}\frac{(p_{0}^{2}-p_{1}^{2})^{2}}{m^{4}}= p_{0}^{2}-p_{1}^{2}-\ell p_{0} p_{1}^{2}.
\eea
which is what we expected from (\ref{eq:MetricandDispersionRelation}).

In the above formulas we left the indication of a possible dependence on coordinates $x$, even if in the particular case we are considering the metric only depends on momenta $p$. This is to emphasize that $g_{\mu\nu}$ is a full-fledged Finsler metric, which would acquire an explicit coordinates dependence in the case of a non-flat spacetime. 
Indeed, the metric (\ref{fmetric}) is in every respect a metric on spacetime, which allows to define an inner product structure over the tangent bundle.
The  dependence on velocity/momentum and mass implies that different particles with different mass and/or velocity "see" different spacetimes.

Despite its dependence on velocity/momentum, this metric should not be confused with the non-trivial momentum space metric that has been quite robustly associated to departures from special-relativistic symmetries (see introduction and end of section \ref{sec:kappaPoincare}).
In section \ref{sec:LineElement} we will elaborate more on the different roles of these two metrics, $g_{\mu\nu}$ and $\zeta^{\mu\nu}$. Here let us just mention the fact that also the momentum-space metric $\zeta^{\mu\nu}$ is related to the particle dispersion relation, but in a less immediate way (\cite{AmelinoCamelia:2011bm, GiuliaFlaviokappaLocality}): the dispersion relation of a particle with momentum $p$ can be obtained by computing the geodesic distance from the origin of momentum space to the point $p$, where the geodesic is the one defined by the metric $\zeta^{\mu\nu}(p)$.

\subsection{Worldlines}
We have shown that it is indeed possible to construct a Finsler geometry starting from a modified dispersion relation that is compatible with $\kappa$-Poincar\'e symmetry group.
Still, the issue of establishing up to which point the Finsler geometry framework can correctly describe the physics of a particle with $\kappa$-Poincar\'e symmetries is not solved:
we still need to check if it allows to correctly infer the particle motion (worldline) and the symmetry transformations under which the worldline is covariant.

In this subsection we will deal with the first issue, constructing  the worldline of a particle living on the Finsler geometry associated to the dispersion relation $m^2=\mathcal C_\ell(p)$.
In the next section we will deal with the issue of symmetry transformations.

Worldlines in Finsler geometry are derived through the Euler-Lagrange equations, which lead to a geodesic equation of the form \cite{MDRFinsler}:
\be
\ddot x^{\mu}+\Gamma^{\mu}_{\nu\rho}(x,\dot x)\dot x^{\nu}\dot x^{\rho}=0, \label{eq:geodesicEQ}
\ee
once one assumes the affine parameterization ($F(x,\dot x)=1$). The (generalized) Christoffel symbol $\Gamma^{\mu}_{\nu\rho}(x,\dot x)$ is defined as a function of derivatives of the metric $g_{\mu\nu}$ with respect to spacetime coordinates in the same way as in Riemannian geometry:
\be
\Gamma^\mu_{\nu\rho}(x,\dot x)=\frac{1}{2}g^{\mu\sigma}(x,\dot x)\left[ -\partial_\sigma g_{\nu\rho}+\partial_\nu g_{\rho\sigma}+\partial_\rho g_{\sigma\nu}\right],
\ee
but it will in general depend on velocities because the metric does.

In the case we are studying, where the metric is given by eq.~(\ref{fmetric}), the associated Christoffel symbols vanish, and the geodesic equation reduces to:
\be
\ddot x^{\mu}=0,\label{eq:FinslerEOM}
\ee
\emph{i.e.} $\dot x^{\mu}=\text{constant}$.
One finds that the affine parameterization\footnote{
Notice that, in analogy to what happens in general relativity ($\ell=0$), there is some freedom in redefining the affine parameter in such a way that  the form of the worldlines is unchanged. In the classical case, $F=\sqrt{(\dot x^0)^2-(\dot x^1)^2}$ and one can choose the affine parameter so to fix $F$ to any real constant, leaving the coordinate velocity invariate. In our case ($\ell \neq 0$), one can choose the affine parameter so that $F(x,\dot x)=C(1+\ell f(\dot{x}))$, with $C$ a real constant and $f$ a generic function of velocities. 
This will lead to a coordinate velocity that differs from the one appearing in (\ref{eq:FinslerWorldline}) only by terms proportional to $\ell C^2 f(\dot x)$ when written in terms of velocities, and it will coincide with it when they are both written in terms of momenta.
} $F=1$ applied to the norm  (\ref{eq:dSFinslerNorm}) implies the following relation between $\dot x^{1}$ and $\dot x^{0}$:
\be
(\dot{x}^0)^2-(\dot{x}^1)^2+\ell m\frac{\dot{x}^0(\dot{x}^1)^2}{(\dot{x}^0)^2-(\dot{x}^1)^2}=1\Rightarrow \;\;(\dot{x}^1)^2=(\dot{x}^0)^2-1+\ell m\dot x^{0}((\dot{x}^0)^2-1).\label{eq:Feq1}
\ee
Upon integration  along the affine parameter one gets:
\be
x^{1}-\bar x^{1}=\left( \frac{\sqrt{(\dot x^{0})^{2}-1}(1+\frac{\ell}{2} m \dot x^{0})}{\dot x^{0}} \right)(x^{0}-\bar x^{0}), \label{eq:FinslerWorldline}
\ee
where $\dot x^{0}$ is constant along the motion because of (\ref{eq:FinslerEOM}).

One can easily verify that this worldline is equivalent to the one found within the $\kappa$-Poincar\'e framework, eq. (\ref{eq:dSworldlineMassive}): it is sufficient to write $\dot x^{0}$ as a function of momenta using (\ref{eq:x0dotVSp}) and taking into account the constraint (\ref{eq:Feq1}). Then using the dispersion relation one writes the resulting function of $p_{0}$ and $p_{1}$ as a function of $p_{0}$ and the mass only. This gives a worldline with the same form as (\ref{eq:dSworldlineMassive}).

So the issue of  describing the particle motion is set: the Finsler framework allows to derive the same worldlines as the ones that are found by applying the Hamiltonian  formalism to the $\kappa$-Poincar\'e phase space (see section \ref{sec:kappaPoincare}).

\section{Symmetries}
\label{sec:symmetries}

We have shown that the same worldlines are obtained in the two frameworks of $\kappa$-Poincar\'e group and Finsler geometry with $\kappa$-Poncar\'e-inspired dispersion relation. In order to fully understand whether the two frameworks are physically equivalent  we also need to compare the symmetry transformations derived within the Finsler framework with the ones generated by the $\kappa$-Poincar\'e group.

\subsection{Covariance of worldlines under $\kappa$-Poincar\'e symmetry transformations}

The $\kappa$-Poincar\'e group generators, whose Lie brackets and representation were given in section \ref{sec:kappaPoincare}, define the symmetries of a free particle with $\kappa$-Poincar\'e inspired dispersion relation in the sense that  the particle worldline is covariant under their action:
\be
(x^{1})^{\prime}=v( p_{0}^{\prime})\cdot (x^{0})^{\prime}\; \Leftrightarrow \;x^{1}=v(p_{0})\cdot  x^{0}, \label{eq:covarianceOFworldline}
\ee
where for simplicity we set the initial condition $\bar x^{0}=\bar x^{1}=0$ and $v(p_{0})$ is the coordinate velocity found in (\ref{eq:dSworldlineMassive}). Given a generic function of coordinates and momenta $f(x,p)$, $f(x,p)^{\prime}$ indicates the transformed form of the function: for a boost transformation,  at the first order in the boost parameter $\xi$, one has $f(x,p)^{\prime}=f(x,p)+\xi \left\{\mathcal N,f(x,p)\right\}$.
The above condition (\ref{eq:covarianceOFworldline}) is in this case equivalent to asking
\be
\left\{\mathcal N,x^{1}\right\}=v(p_{0})\left\{\mathcal N,x^{0}\right\}+\frac{\partial v(p_{0})}{\partial p_{0}}\left\{\mathcal N,p^{0}\right\}x^{0} \; \Leftrightarrow \;x^{1}=v(p_{0}) \;x^{0},
\ee
which is easily verified by making use of the Lie brackets (\ref{eq:kappaAlgebra}) and (\ref{eq:BoostCoordParentheses}) and of the dispersion relation.

\subsection{Finsler Killing vectors}

Finsler geometry provides us with a prescription for deriving the symmetries of the metric, which relies on  the Killing equations  \cite{MDRFinsler,HRund}  (see appendix \ref{sec:DerivationOfKilling} for a detailed derivation):
\begin{equation}
g_{\mu\rho} \, \partial_\nu \xi^\rho +
g_{\nu\rho} \, \partial_\mu \xi^\rho+
\frac{\partial g_{\mu\nu}}{\partial\dot{x}^\rho} \, \frac{\partial \xi^\rho}{\partial x^\sigma}\dot{x}^\sigma +\frac{\partial g_{\mu\nu}}{\partial x^\rho}\xi_\rho= 0.
\label{eq:Killing1}
\end{equation}
Notice that in the case we are studying, where the metric is given by eq.(\ref{fmetric}), the last term in the equation does not contribute since the metric is independent of coordinates.
We solve this differential equation looking for a perturbative solution at first order in $\ell$:
\begin{equation}
\xi^{\mu}=\xi_{(0)}^{\mu} + \ell \, \xi_{(1)}^{\mu}.
\end{equation}
The zeroth order  is given by the standard Minkowski spacetime Killing vector:
\begin{equation}
\xi_{(0)}^{\mu}=
\begin{pmatrix}
a\, x^1 + d^0 \\
a\, x^0 + d^1
\end{pmatrix},\label{eq:xi0}
\end{equation}
where $a$, $d^0$ and $d^1$ are the parameters associated, respectively,
to boost, time and spatial translation.
The first-order part of the solution is:
\begin{equation}
\xi_{(1)}^{\mu}=
\begin{pmatrix}
A^0+d^0 m F_{[1]}(\dot{x})+C x^1 + a m (F_{[2]}(\dot{x}) x^0 + F_{[3]}(\dot{x}) x^1)\\
&\\
A^1+d^1 m F_{[4]}(\dot{x})+C x^0 + a m (F_{[5]}(\dot{x}) x^0 + F_{[6]}(\dot{x}) x^1)
\end{pmatrix},
\label{eq:xi1}
\end{equation}
where we have defined:
\begin{eqnarray}
&& F_{[1]}(\dot{x}) = \frac{(\dot{x}^1)^2 ((\dot{x}^0)^2+(\dot{x}^1)^2)}{2 \dot{x}^0 ((\dot{x}^0)^2-(\dot{x}^1)^2)^{3/2}},\\
&& F_{[2]}(\dot{x}) = \frac{(\dot{x}^1)^2 (4(\dot{x}^0)^2\dot{x}^1+5(\dot{x}^1)^3)}{4 ((\dot{x}^0)^2-(\dot{x}^1)^2)^{5/2}},\\
&& F_{[3]}(\dot{x}) = \frac{(\dot{x}^1)^2 (-14(\dot{x}^0)^3+5\dot{x}^0(\dot{x}^1)^2)}{4 ((\dot{x}^0)^2-(\dot{x}^1)^2)^{5/2}},\\
&& F_{[4]}(\dot{x}) = \frac{(\dot{x}^1)^3}{((\dot{x}^0)^2-(\dot{x}^1)^2)^{3/2}},\\
&& F_{[5]}(\dot{x}) = \frac{4 (\dot{x}^0)^5-8 (\dot{x}^0)^3(\dot{x}^1)^2-5\dot{x}^0 (\dot{x}^1)^4}{4 ((\dot{x}^0)^2-(\dot{x}^1)^2)^{5/2}},\\
&& F_{[6]}(\dot{x}) = \frac{12 (\dot{x}^0)^4 \dot{x}^1-5 (\dot{x}^0)^2(\dot{x}^1)^3+2 (\dot{x}^1)^5}{4 ((\dot{x}^0)^2-(\dot{x}^1)^2)^{5/2}},
\end{eqnarray}
and the integration constants $A^{0},\; A^{1}$ and $C$ can be in general functions of the velocities and the mass $m$, with $A^0, \, A^1$ dimensionless and $C$ having the dimension of a mass.

The family of  Killing vectors associated to the metric $g_{\mu\nu}$ has thus more degrees of freedom than the ones associated to the usual (Riemannian, maximally symmetric, 1+1D) three-free-parameters symmetries: it contains three free parameters and three free functions of the velocities. At the end of this section we will elaborate about the role of these additional degrees of freedom, but first we will show that we indeed recover the result known from the study of the $\kappa$-Poincar\'e group, \emph{i.e.} the Killing vectors derived within the Finsler geometry framework do reproduce the symmetries described by the $\kappa$-Poincar\'e generators.

It will turn useful to write down explicitly the conserved charges associated to the Killing vectors
\begin{equation}
Q_{F}=\xi^\mu \, p_\mu(\dot{x}).
\label{charge}
\end{equation}
Writing the charge perturbatively as $Q_{F}=Q_{F}^{(0)}+\ell Q_{F}^{(1)}$ one has
\bea
Q_{F}^{(0)}&=&m \frac{d^{0}\dot x^{0}-d^{1}\dot x^{1}-a x^{0}\dot x^{1}+a x^{1}\dot x^{0}}{\sqrt{(\dot{x}^0)^2 - (\dot{x}^1)^2}},\nonumber\\
Q_{F}^{(1)}&=& m \left[\frac{A^{0}\dot x^{0}-A^{1}\dot x^{1}}{\sqrt{(\dot x^{0})^{2}-(\dot x^{1})^{2}}}+\frac{C (\dot x^{0}x^{1}-\dot x^{1}x^{0})}{\sqrt{(\dot x^{0})^{2}-(\dot x^{1})^{2}}}\right].\label{eq:FinslerCharge}
\eea
If one makes use of the relation between velocities and momenta that was given in eqs. (\ref{eq:x0dotVSp}) and (\ref{eq:x1dotVSp}) it is possible to write the above family of conserved charges in terms of momenta
\bea
Q_{F}^{(0)}&=&d^{0}p_{0}+d^{1}p_{1}+a x^{0}p_{1}+a x^{1}p_{0}\\
Q_{F}^{(1)} & = &
A^0 p_0+A^1 p_1+C (p_0 x^1+p_1 x^0)+\nonumber\\
&& \frac{a \left(2 p_0^3 p_1 x^0+p_1^2 x^1 \left(p_0^2+p_1^2\right)\right)}{2
   m^2}+\frac{d^0 p_1^2 \left(p_0^2+p_1^2\right)}{2 m^2}+\frac{d^1 p_0^3 p_1}{m^2},
\eea
(but note that $C, A^{0}, A^{1}$ remain free functions of velocities).

\subsection{Comparison between Killing vectors and $\kappa$-Poincar\'e symmetries}
\label{sub:comparisonWkappa}
As we mentioned, we want to compare the symmetry transformations derived from  the Killing equation in the Finsler framework and the symmetries generated by $\kappa$-Poincar\'e group.

Since the Killing-Finsler symmetries appear to have more degrees of freedom than the $\kappa$-Poincar\'e ones (three parameters and three functions versus three parameters only), one could think that they describe transformations that are more general than the ones of $\kappa$-Poincar\'e.
Indeed, we are going to show that the Killing-Finsler symmetries contain as a special case the ones generated by the $\kappa$-Poincar\'e generators appearing in section \ref{sec:kappaPoincare}.
To this aim we will compare the conserved charges derived in the two frameworks, so it is convenient to write down the $\kappa$-Poincar\'e conserved charges - obtained from the representation of the symmetry generators in the phase space - in terms of the velocities.

We report here for convenience the representation of the boost generator (the one of translation generators is trivial):
\begin{equation}
\mathcal{N}_{\kappa}(x,p) = p_1 x^0 + p_0 x^1 - x^1 \left(p_0^2 + \frac{p_{1}^2}{2}\right) \ell.\label{eq:kappaBoostBicross}
\end{equation}
Thanks to the relation between momenta and velocities provided by eqs. (\ref{eq:p0_complete}) and (\ref{eq:p1_complete}) this becomes:
\begin{equation}
\mathcal{N}_{\kappa}(x,\dot x) = m \left(\frac{\dot{x}^0 x^1 - \dot{x}^1 x^0}{\sqrt{(\dot{x}^0)^2 - (\dot{x}^1)^2}} +
   m \, (\dot{x}^0)^3 \frac{\dot{x}^1 x^0 - \dot{x}^0 x^1}{((\dot{x}^0)^2 - (\dot{x}^1)^2)^2}\, \ell\right). \label{eq:Nkappavsxdot}
\end{equation}

A generic $\kappa$-Poincar\'e transformation is a combination of a boost, a time translation and a space translation, and so the generic charge is:
\be
Q_{\kappa}=\mathsf{A} \mathcal N_{\kappa}+\mathsf{B}\mathcal P_{0}+\mathsf{C} \mathcal P_{1}, \label{eq:Qkappa}
\ee
where $\mathsf{A},\mathsf{B},\mathsf{C}$ are the transformation parameters. Using (\ref{eq:Nkappavsxdot}), (\ref{eq:p0_complete}) and (\ref{eq:p1_complete}) one can then write the generic $\kappa$-Poincar\'e charge in terms of velocities:
\bea
Q_{\kappa}&=&\mathsf{A} \, m\left[ \left(\frac{\dot{x}^0 x^1 - \dot{x}^1 x^0}{\sqrt{(\dot{x}^0)^2 - (\dot{x}^1)^2}} +
   m \, (\dot{x}^0)^3 \frac{\dot{x}^1 x^0 - \dot{x}^0 x^1}{((\dot{x}^0)^2 - (\dot{x}^1)^2)^2}\, \ell\right)\right]\nonumber\\&&+\mathsf{B} \, m\left[\frac{\dot x^{0}}{\sqrt{(\dot x^{0})^{2}-(\dot x^{1})^{2}}} -\ell m \frac{(\dot x^{1})^{2}((\dot x^{0})^{2}+(\dot x^{1})^{2})}{2((\dot x^{0})^{2}-(\dot x^{1})^{2})^{2}}\right]\nonumber\\
   &&+\mathsf{C}  \, m\left[-\frac{\dot x^{1}}{\sqrt{(\dot x^{0})^{2}-(\dot x^{1})^{2}}} +\ell m \frac{\dot x^{1}(\dot x^{0})^{3}}{((\dot x^{0})^{2}-(\dot x^{1})^{2})^{2}}\right].
\eea
In order to verify that the Finsler family of charges given in eq.(\ref{eq:FinslerCharge}) contains the $\kappa$- Poincar\'e one as special case, we ask  the two charges $Q_{\kappa}$ and $Q_{F}$ to be equal.
At zero order in $\ell$ this implies
\be
\mathsf{A}=a, \quad \mathsf{B}=d^{0},\quad \mathsf{C}=d^{1}.
\ee
Introducing this into the first-order terms and comparing them one gets a set of constraints on the functions $A^0,\, A^1$ and $C$, which read
\begin{equation}
A^1 =
-d^1 \, m \frac{(\dot{x}^0)^3}{((\dot{x}^0)^2-(\dot{x}^1)^2)^{3/2}}
+d^0 \, m \frac{\dot{x}^1 ((\dot{x}^0)^2+(\dot{x}^1)^2)}{((\dot{x}^0)^2-(\dot{x}^1)^2)^{3/2}}
+A^0 \frac{\dot{x}^0}{\dot{x}^1},
\label{kCondition1}
\end{equation}
and
\begin{equation}
C=-a \, m \frac{(\dot{x}^0)^3}{((\dot{x}^0)^2-(\dot{x}^1)^2)^{3/2}}.
\label{kCondition2}
\end{equation}
One can verify that this conditions are compatible with the Killing vector $\xi$  to be still a solution of the Killing equation (\ref{eq:Killing1}).

So we can conclude that there exists one choice of the free parameters and functions in the Killing-Finsler symmetries that reproduces the $\kappa$-Poincar\'e ones.
The issue is now to understand  what kind of transformations are described by other choices of the free functions.
In the following subsection we show that the additional freedom provided by the free functions mirrors the freedom that one has to redefine the $\kappa$-Poincar\'e symmetry generators, in such a way that the invariance of the Casimir is preserved under the new transformations.

\subsection{Generators redefinition in $\kappa$-Poincar\'e and free functions in Finsler Killing vectors}

In $\kappa$-Poincar\'e one can redefine the boost by performing a diffeomorphism, so that the Casimir is still invariant under the action of the new generator \footnote{This redefinition will of course require to coherently modify the Lie brackets relations between the boost and translations generators.}. We will show that the freedom provided by the free functions appearing in the Finsler conserved charges is actually the freedom needed to span the possible redefinitions of the boost generator.

To do this, we work at the level of the the boost representation in coordinates and momenta. One can consider the most generic deformation of the classical boost by adding to the classical boost all possible corrections at the first order in $\ell$. The allowed terms, from a dimensional point of view, are monomial of the form $\ell x^{\mu}\,p_{\nu}p_{\alpha}$ or $\ell p_{\alpha}$
\be
\mathcal N_{generic}= p_1 x^0 + p_0 x^1+\ell\left(X p_{0}+Y p_{1}+ \alpha p_{0}p_{1}x^{0}+\beta p_{0}^{2}x^{0}+\gamma p_{0}p_{1}x^{1}+\delta p_{0}^{2}x^{1}+\zeta p_{1}^{2}x^{0}+\eta p_{1}^{2}x^{1} \right)
\label{Generic_Boost}
\ee
where $X,Y,\alpha..\eta$ are numerical coefficients.

The condition that has to be satisfied by the new boost in order for it to be considered a legitimate redefinition of the $\kappa$-Poincar\'e one is that it has vanishing Lie brackets with the Casimir (so that it still describes a symmetry of the system):
\be
\left\{\mathcal N_{generic},C_{\ell}\right\}=0.
\ee
This imposes some constraints on the parameters:
\bea
&&\alpha-\delta=1 \nonumber\\
&&\eta=-\frac{1}{2}\nonumber\\
&&\beta=0 \label{eq:conditionsforcompatibility}\\
&&\gamma=\zeta \nonumber
\eea
so that at the end we have  two  free parameters governing the possible combinations of the monomials $\ell x^{\mu}\,p_{\nu}p_{\alpha}$ that can appear in the new boost, plus the parameters $X$ and $Y$ that multiply "translation-like" terms:
\be
\mathcal N_{\kappa-compatible}= p_1 x^0 + p_0 x^1+\ell\left(X p_{0}+Y p_{1}+ \alpha p_{0}p_{1}x^{0}+\gamma (p_{0}p_{1}x^{1}+ p_{1}^{2}x^{0})+(\alpha-1) p_{0}^{2}x^{1}-\frac{1}{2} p_{1}^{2}x^{1} \right) \label{eq:NkappaCompatible}
\ee

We want to show that the freedom of redefining the boost generator without spoiling the invariance of the $\kappa$-Poincar\'e Casimir is related to the presence of the free functions in the Finsler charge.
To this aim,  we compare the Finsler charge $Q_{F}$ with the generic family of charges obtained from $\mathcal N_{generic}$ and the translations:
\be
Q_{generic}=\mathsf{A} \mathcal N_{generic}+\mathsf{B}\mathcal P_{0}+\mathsf{C} \mathcal P_{1},
\ee
in an analogous way to what was done at the end of the previous subsection (see eqs. (\ref{eq:Qkappa})-(\ref{kCondition2})). Notice that this charge, $Q_{generic}$, is not in general a conserved charge, since we didn't impose the constraints (\ref{eq:conditionsforcompatibility}) on the free parameters it contains.

Comparing $Q_{F}$ and $Q_{generic}$ we observe that the two parameters $X$ and $Y$ that appear in $\mathcal N_{generic}$ multiply the same kind of terms that are multiplied by $A^{0}$ and $A^{1}$ in the charge $Q_{F}$: the presence of $A^0$ and $A^1$ in $Q_F$ is due to the freedom of adding "translation-like" terms to the boost generator.
Since the issue about the $A^0$ and $A^1$ terms in $Q_F$ is solved thanks to the $X$ and $Y$ terms in $Q_{generic}$, we now focus only on the terms multiplying greek-letter parameters in $Q_{generic}$ and the $a$ and $C$ terms in $Q_{F}$. So we ask that $Q_{generic}\Big|_{\mathsf{B}=\mathsf{C}=X=Y=0}=Q_{F}\Big|_{d^{0}=d^{1}=A^{0}=A^{1}=0}$.

\vspace{3mm}
It turns out that it is possible to match the two charges only if the greek-letter parameters satisfy exactly the constraints (\ref{eq:conditionsforcompatibility}).
Given these constraints, the matching is achieved for
\bea
C&=&m  \frac{\gamma\left((\dot x^1)^3-(\dot x^0)^2\dot x^1\right)+\delta\left((\dot x^0)^3-\dot x^0(\dot x^1)^2\right)-\dot x^0 (\dot x^1)^2}{((\dot x^{0})^{2}-(\dot x^{1})^{2})^{3/2}}.
\eea
So  the only admissible form of the boost sector of the Finsler charge is the one that is compatible with the $\kappa$-Poincar\'e Casimir and that is linked to the boost in  bicrossproduct basis by a diffeomorphism: the  freedom provided by the free functions of velocity in the Killing-Finsler charge corresponds to the freedom that we have to redefine the boost in such a way that it still leaves  the Casimir invariant.

\section{Finsler geometry of another $\kappa$-Poincar\'{e} basis}

In the previous sections we have studied the Finsler geometry associated to the Casimir of $\kappa$-Poincar\'e algebra in the so-called bicrossproduct basis.

We have found that the Finsler formalism leads to equivalent results as far as the worldlines are concerned. The associated symmetries are also compatible with the ones of $\kappa$-Poincar\'e, and in particular the symmetries derived within the Finsler formalism 'know' about the possibility of redefining the boost generator of $\kappa$-Poincar\'e leaving the Casimir invariant.

In this section we investigate what happens upon a (nonlinear) redefinition of the translation generators in the $\kappa$-Poincar\'e algebra. This is an allowed redefinition within the formalism of Hopf algebras.  Of course a redefinition of the translation generators requires that one changes the Casimir accordingly.

Since the Casimir is modified, the associated Finsler geometry will be different, and one might wonder if also in this case Finsler geometry allows to  reproduce the features of $\kappa$-Poincar\'e (the ones that are proper of this basis). In particular the issue is whether one recovers the correct form of conserved charges, reproducing the representation of $\kappa$-Poincar\'e in this basis. We are going to show that this is indeed the case.

We choose the basis of $\kappa$-Poincar\'e that has the same Casimir as the one studied in \cite{MDRFinsler}:
\be
\mathcal C_{\ell}^{(new)}=p_{0}^{2}-p_{1}^{2}-\ell p_{1}^{3}.
\label{LiberatiMDR_casimir}
\ee

This new Casimir  has the peculiarity to be non isotropic: it contains a term which is odd in the spatial momentum ($\ell p_1^3$). We used this peculiar Casimir for two reasons: it appeared before in the literature; we needed a new Casimir to test our approach.
If one wants to obtain an isotropic Casimir, one can substitute the odd term with, for example ($\ell |p_1|^3$). We choose to use the odd one for the sake of simplicity in the calculations.

There is a diffeomorphism in the space of the \emph{generators} of the algebra that connects the bicrossproduct basis with this one, such that the Casimir (\ref{eq:Casimirk}) becomes (\ref{LiberatiMDR_casimir})\footnote{Notice that the diffeomorphisims that we are performing here are transformations in the momentum space and not in spacetime.}:
\bea
&& p_0 \rightarrow p_0, \nonumber\\
&& p_1 \rightarrow p_1 \left(1+\frac{\ell}{2}(p_1 - p_0)\right).
\label{diffeo}
\eea
The resulting algebra is the following
\bea
&&\{\mathcal{N},p_0\} = p_1 \left(1+\frac{\ell}{2}(p_1 - p_0)\right),\nonumber\\
&&\{\mathcal{N},p_1\} = p_0 \left(1-\frac{\ell}{2} p_0 -\ell p_1\right). \label{eq:algebraNewBasis}
\eea

The corresponding representation (choosing the ordinary symplectic structure) of the boost is
\be
\mathcal{N}=p_1 x^0 + p_0 x^1 - \ell \left(\frac{1}{2} p_0 p_1 x^0 + p_0 p_1 x^1 + \frac{1}{2} p_0^2 x^1 - \frac{p_1^2 x^0}{2}\right).    \label{eq:BoostRepNew}
\ee

Note that the prescription given in~\cite{MDRFinsler} assumes only a Casimir as input to obtain the Finsler model. Applying naively the momenta redefinition (\ref{diffeo}) to the action (\ref{eq:kappaAction}) would lead to additional terms proportional to $\dot{x}^\mu$, corresponding to non trivial symplectic structure.
Our aim is to compare two \emph{different} Finsler models and to study the relation between the two, given a map between them on the quantum group side. Changing the symplectic structure would correspond to a passive transformation, which would only give trivial results.

The  Finsler norm associated to the Casimir (\ref{LiberatiMDR_casimir}) was found already in \cite{MDRFinsler}:

\be
F^{(new)}=\sqrt{(\dot x^{0})^{2}-(\dot x^{1})^{2}}-\frac{\ell}{2}m \frac{(\dot x^{1})^{3}}{(\dot x^{0})^{2}-(\dot x^{1})^{2}}
\ee

From the norm one gets the metric

\be
g_{\mu\nu}^{(new)}=
\begin{pmatrix}
1-\ell m\, (\dot x^{1})^{3}\frac{(\dot x^{0})^{2}+\frac{1}{2} (\dot x^{1})^{2}}{((\dot x^{0})^{2}-(\dot x^{1})^{2})^{\frac{5}{2}}} & \frac{3}{2} \ell m\frac{(\dot x^{0})^{3}(\dot x^{1})^{2}}{((\dot x^{0})^{2}-(\dot x^{1})^{2})^{\frac{5}{2}}} \\
 \frac{3}{2}\ell m\frac{(\dot x^{0})^{3}(\dot x^{1})^{2}}{((\dot x^{0})^{2}-(\dot x^{1})^{2})^{\frac{5}{2}}} & -1-\ell m\, \dot x^{1}\frac{3(\dot x^{0})^{4}-\frac{5}{2}(\dot x^{0})^{2}(\dot x^{1})^{2}+(\dot x^{1})^{4}}{((\dot x^{0})^{2}-(\dot x^{1})^{2})^{\frac{5}{2}}}
\end{pmatrix}
\ee
The relation between momenta and velocities is
\bea
p_{0}(\dot x)&=& m \frac{\dot x^{0}}{\sqrt{(\dot x^{0})^{2}-(\dot x^{1})^{2}}}\left(1+\ell m\frac{ (\dot x^{1})^{3}}{((\dot x^{0})^{2}-(\dot x^{1})^{2})^{\frac{3}{2}}}\right) \label{eq:p0_otherbasis}\\
p_{1}(\dot x)&=& - m \frac{\dot x^{1}}{\sqrt{(\dot x^{0})^{2}-(\dot x^{1})^{2}}}\left(1- \frac{1}{2}\ell m \dot x^{1} \frac{(\dot x^{1})^{2} -3 (\dot x^{0})^{2}}{((\dot x^{0})^{2}-(\dot x^{1})^{2})^{\frac{3}{2}}}\right) \label{eq:p1_otherbasis}
\eea

The Killing vectors of this metric are:
\begin{equation}
\xi^{\mu}=\xi_{(0)}^{\mu} + \ell \, \xi_{(1)}^{\mu}
\end{equation}
With the zeroth order  given as before by the Minkowski spacetime Killing vector:
\begin{equation}
\xi_{(0)}^{\mu}=
\begin{pmatrix}
a\, x^1 + d^0 \\
a\, x^0 + d^1
\end{pmatrix}
\end{equation}
The first-order part of the Killing vectors is:
\begin{equation}
\xi_{(1)}^{\mu}=
\begin{pmatrix}
x^1 D +a\, m \, (\dot{x}^1)^3 \, (x^0 \, G_{[1]}(\dot{x})+x^{1}\, G_{[2]}(\dot x))
+B^0 \\
&&\\
x^0 D -
   a\, m\,( \dot{x}^1)^{3}\, x^1 G_{[3]}(\dot{x})+B^1
\end{pmatrix}
\end{equation}
where $D,\,B^{1},\, B^{0}$ are free functions of velocities and
\bea
G_{[1]}(\dot{x})&=&\frac{9}{4}\frac{\dot x^{0}\dot x^{1}}{((\dot x^{0})^{2}-(\dot x^{1})^{2})^{\frac{5}{2}}}\\
G_{[2]}(\dot{x})&=&\frac{3}{2}\frac{-4 (\dot x^{0})^{2}+(\dot x^{1})^{2}}{((\dot x^{0})^{2}-(\dot x^{1})^{2})^{\frac{5}{2}}}\\
G_{[3]}(\dot{x})&=&\frac{3}{4}\frac{2 (\dot x^{0})^{2}+ (\dot x^{1} )^2}{((\dot x^{0})^{2}-(\dot x^{1})^{2})^{\frac{5}{2}}}
\eea
The conserved charges associated to the family of Killing vectors are found as:
\be
Q^{(new)}=p_\nu(\dot x)\xi^\nu=\frac{m}{F^{(new)}}g_{\mu\nu}^{(new)}\dot x^{\mu}\xi^{\nu}
\ee
Note that the formal expression of the charge is the same as the one used before, in section \ref{sec:symmetries}, but  the norm and the metric (and of course the Killing vectors) are different functions of the velocities than the ones of section \ref{sec:symmetries}.

The resulting family of charges is
\bea
Q^{(new)}&=& m \frac{d^{0}\dot x^{0}-d^{1}\dot x^{1}-a x^{0}\dot x^{1}+a x^{1}\dot x^{0}}{\sqrt{(\dot{x}^0)^2 - (\dot{x}^1)^2}}\nonumber\\
&&+ \ell m\Bigg[D \frac{\dot x^{0} x^{1}-\dot x^{1}x^{0}}{\sqrt{(\dot x^{0})^{2}-(\dot x^{1})^{2}}}+\frac{\dot x^{0}B^{0}-\dot x^{1} B^{1}}{(\dot x^{0})^{2}-(\dot x^{1})^{2}}\nonumber\\
&&\qquad\quad+m (\dot x^{1})^{2}\frac{d^{0}\dot x^{0}\dot x^{1}\left((\dot x^{0})^{2}-(\dot x^{1})^{2}\right)-\frac{1}{2} d^{1}\left(3 (\dot x^{0})^{2}-4(\dot x^{0})^{2}(\dot x^{1})^{2}+(\dot x^{1})^{4}\right)}{\left((\dot x^{0})^{2}-(\dot x^{1})^{2}\right)^{3}}\nonumber\\
&&\qquad\quad+\frac{ a \,m \dot x^{1}}{4 \left((\dot x^{0})^{2}-(\dot x^{1})^{2}\right)^{3}}\left(6 (\dot x^{0})^{4}-17 (\dot x^{0})^{2}(\dot x^{1})^{2}+2 (\dot x^{1})^{4}\right)(\dot x^{0}x^{1}-\dot x^{1}x^{0})\Bigg] \label{eq:newCharge}
\eea

We verify that this charge cannot reproduce the conserved charges of $\kappa$-Poincar\'e in the bicrossproduct basis. Then we will show that is does instead reproduce the conserved charges in the appropriate basis (\ref{eq:algebraNewBasis}).

Concerning the comparison with the bicrossproduct basis, we proceed as in subsection \ref{sub:comparisonWkappa}. The only difference is the relation between momenta and velocities that we have to use in order to re-write the representation of boosts (\ref{eq:kappaBoostBicross}) and translations in terms of velocities. In fact now the relations to be used are (\ref{eq:p0_otherbasis}) and (\ref{eq:p1_otherbasis}) instead of (\ref{eq:p0_complete}) and (\ref{eq:p1_complete}).

Upon doing this one gets the bicrossproduct-basis charge, represented in the new velocity space
\bea
Q_{\kappa}&=&d^{0}\left( \frac{m \dot x^{0}}{\sqrt{(\dot{x}^0)^2 - (\dot{x}^1)^2}}+\ell \frac{m^{2}\dot x^{0} (\dot x^{1})^{3}}{\left((\dot{x}^0)^2 - (\dot{x}^1)^2\right)^{2}} \right)+d^{1}\left(- \frac{m \dot x^{1}}{\sqrt{(\dot{x}^0)^2 - (\dot{x}^1)^2}}+\ell m^{2}(\dot x^{1})^{2} \frac{(\dot x^{1})^{2}-3 (\dot x^{0})^{2}}{2\left((\dot{x}^0)^2 - (\dot{x}^1)^2\right)^{2}} \right)\nonumber\\
&&+a \Bigg[ \frac{m (\dot x^{0}x^{1}-\dot x^{1}x^{0})}{\sqrt{(\dot{x}^0)^2 - (\dot{x}^1)^2}}+ \ell m^{2} \frac{x^{1}\left( (\dot x^{1})^{4}+2 \dot x^{0}(\dot x^{1})^{3} +(\dot x^{0})^{2}(\dot x^{1})^{2}-2 (\dot x^{0})^{4}\right)+x^{0}\left((\dot x^{1})^{4}-3 (\dot x^{0})^{2}(\dot x^{1})^{2}\right)}{2 \left((\dot x^{0})^{2}-(\dot x^{1})^{2}\right)^{2}}  \Bigg]\nonumber
\eea

It is easy to check that there is no choice of the velocity-dependent functions $D, B^{0}, B^{1}$ such that the charge (\ref{eq:newCharge}) takes this form (one would have to ask these functions to depend on coordinates as well in order to find a map between the two families of charges, but this is incompatible with the Killing equations).

On the other hand, the conserved charge (\ref{eq:newCharge}) reproduces the conserved charge of $\kappa$-Poincar\'e in the basis of momenta that is compatible with the Casimir we are considering. This charge is obtained by using the boost representation (\ref{eq:BoostRepNew}) and mapping it (as well as the momenta themselves) to the velocity space using (\ref{eq:p0_otherbasis}) and (\ref{eq:p1_otherbasis})
\bea
Q^{\prime}_{\kappa}&=&d^{0}\left( \frac{m \dot x^{0}}{\sqrt{(\dot{x}^0)^2 - (\dot{x}^1)^2}}+\ell \frac{m^{2}\dot x^{0} (\dot x^{1})^{3}}{\left((\dot{x}^0)^2 - (\dot{x}^1)^2\right)^{2}} \right)+d^{1}\left(- \frac{m \dot x^{1}}{\sqrt{(\dot{x}^0)^2 - (\dot{x}^1)^2}}+\ell m^{2}(\dot x^{1})^{2} \frac{(\dot x^{1})^{2}-3 (\dot x^{0})^{2}}{2\left((\dot{x}^0)^2 - (\dot{x}^1)^2\right)^{2}} \right)\nonumber\\
&&+ a\, m\frac{ \dot x^{0}x^{1}-\dot x^{1}x^{0}}{\sqrt{(\dot{x}^0)^2 - (\dot{x}^1)^2}} \Bigg[ 1+ \frac{1}{2} \ell m \dot x^{0} \frac{(\dot{x}^1)^2-(\dot{x}^0)^2+2 \dot x^{0} \dot x^{1}}{ \left((\dot x^{0})^{2}-(\dot x^{1})^{2}\right)^{3/2}}  \Bigg]\nonumber
\eea

Asking that this charge is reproduced by the Finsler one, eq.~(\ref{eq:newCharge}), amounts to fix the free functions $D,B^{0},B^{1}$ in the following way:
\bea
D&=&- a \, m \frac{2 (\dot x^{0})^{5} +2 (\dot x^{0})^{4} \dot x^{1} -4 (\dot x^{0})^{3}(\dot x^{1})^{2} -13 (\dot x^{0})^{2}(\dot x^{1})^{3}+2 \dot x^{0}(\dot x^{1})^{4} +2 (\dot x^{1})^{5}}{4 \left((\dot x^{0})^{2}-(\dot x^{1})^{2}\right)^{5/2} }\nonumber\\
B^{0}&=&B^{1} \frac{\dot x^{1}}{\dot x^{0}}
\eea

As seen already in subsection \ref{sub:comparisonWkappa}, in the case of bicrossproduct basis, it turns out that also in this case the  freedom provided by the free functions of velocity in (\ref{eq:newCharge}) corresponds to the freedom that we have to redefine the boost in such a way that it still leaves invariant the Casimir (\ref{LiberatiMDR_casimir}).

Similarly  to what we did in section \ref{sub:comparisonWkappa}, we write the most generic form of the boost generator, which is again the one in (\ref{Generic_Boost}) and has a number of free parameters $\{\alpha,...,\zeta\}$. The requirement that  the boost is compatible with the Casimir  (\ref{LiberatiMDR_casimir}) translates into the following conditions on the free parameters:
\bea
&& \beta = 0 \label{eq:conditionGenericBoostCompatibility1}\\
&& \alpha - \delta = 0; \\
&& \gamma - \zeta = \frac{3}{2} \\
&& \eta = 0 \label{eq:conditionGenericBoostCompatibility4}
\eea

We then ask under which conditions (on the $\{\alpha,...,\zeta\}$ parameters) it is possible to reproduce the boost-like charge with the Finsler charge  (\ref{eq:newCharge}) for some choice of the free functions of velocities $D, B^{0}, B^{1}$.  And again, as it happened in the case studied before, one has that it is indeed possible to find the correspondence whenever the conditions (\ref{eq:conditionGenericBoostCompatibility1})-(\ref{eq:conditionGenericBoostCompatibility4}) are satisfied.

\section{On the invariance of the Lagrangian}
\label{sec:LineElement}

Let us now discuss about a potential problem formerly stressed in \cite{MignemiFinsler}, about the invariance of the Lagrangian linked to the Finsler norm. The formalism we have described in this paper permits to develop a framework which is consistent with deformed relativistic symmetries: in particular it allows to derive equations of motion which are covariant with respect to these symmetries. One may then be tempted to define the line element of the (Finslerian) spacetime here constructed as $d\sigma={\cal L}d\tau$, with $\cal L$ given by (\ref{eq:dSFinslerLagrangian}), but this quantity is not invariant under deformed boost:
\begin{equation}
\delta_{\cal N}{\cal L}=\{{\cal N},{\cal L}\}=m\frac{\partial F}{\partial \dot{x}^\mu}\{{\cal N},\dot{x}^\mu\}=m\frac{g_{\mu\nu}\dot{x}^\nu}{F}\{{\cal N},\dot{x}^\mu\}=p_\mu\delta_{\cal N}\dot{x}^\mu. \label{pdeltaxpunto}
\end{equation}
The last quantity in the above equation is generally non zero, as it can be shown using equations (\ref{eq:BoostCoordParentheses}), (\ref{eq:dotx0ofp}) and (\ref{eq:dotx1ofp}):
\begin{equation}
p_\mu\delta_{\cal N}\dot{x}^\mu = p_\mu\frac{d}{d\tau}\{\mathcal{N},x^\mu\} = \frac{\ell}{2} \dot{x}^1\left(p_1^2 +2 p_0^2\right)\,.\label{pdeltaxpunto2}
\end{equation}

The fact that the Lagrangian of our theory is not invariant under deformed boost is not however a problem, since under these deformed transformations the Lagrangian only changes by a total derivative:
\begin{equation}
p_\mu\delta_{\cal N}\dot{x}^\mu=\frac{d}{d\tau}(p_\mu \delta_{\cal N} x^\mu)=\frac{d}{d\tau}\left(\frac{\partial{\cal L}}{\partial \dot{x}^\mu}\delta_{\cal N} x^\mu\right)\,,
\end{equation}
(notice that we used the fact that $\dot{p}_\mu = 0$, which follows from the independence of the Lagrangian from coordinates $x^\alpha$).
This is why, despite the non-invariance of the Lagrangian, the action is still invariant and we are still able to derive from it (using standard Euler-Lagrange formalism) covariant equations of motion and  wordlines.
So from a physical point of view the theory we are studying is relativistic with respect to the deformed symmetries we have considered.

The only issue that could be raised by the fact that the Lagrangian is not invariant is the one mentioned at the beginning of this section, concerning the definition of a line element: we cannot define an invariant line element as $ds={\cal L}d\tau$ in analogy with special relativity.
However we will show that it is possible to redefine the Lagrangian in such a way that it  allows for the construction of an invariant line element in the standard way.
Indeed, it is sufficient to add to the Lagrangian $\mathcal L$ terms whose variation under a deformed boost transformation is the same total derivative as the one generated by the Lagrangian.
First of all let us notice that
\begin{equation}
\left\{{\cal N},\frac{\ell}{2}\frac{m^2 \dot{x}^0}{(\dot{x}^0)^2-(\dot{x}^1)^2}(\dot{x}^1)^2\right\}=-\frac{\ell}{2} \dot{x}^1\left(p_1^2 +2 p_0^2\right)\,.\label{boostonedge}
\end{equation}
Consequently, if we recall the definitions of $\lambda(\dot{x})$ parameter (\ref{lambda}) and Finsler norm $F(\dot{x})$, we can realize that our boundary terms are generated by
\begin{equation}
m\left\{{\cal N},2m\lambda(\dot{x}) - F(\dot{x})\right\}=\left\{{\cal N},\frac{\ell}{2}\frac{m^2 \dot{x}^0}{(\dot{x}^0)^2-(\dot{x}^1)^2}(\dot{x}^1)^2\right\} = -p_\mu\delta_{\cal N}\dot{x}^\mu \,.\label{puzzle1}
\end{equation}
Before we move on, let us notice that (\ref{puzzle1}) tells us that $\lambda(\dot{x})$ parameter is invariant under boost transformations $\{{\cal N},\lambda(\dot{x})\}=0$,\footnote{Another possibly interesting result is that, given equations (\ref{eq:dotx0ofp}) and (\ref{eq:dotx1ofp}), now that we know that $\lambda$ parameter is invariant, we can find the relations
\begin{eqnarray}
\left\{{\cal N},\dot{x}^0\right\}&=&\lambda(\dot{x})\left\{{\cal N},2 p_0-\ell p_1^2\right\}=-\dot{x}^1\,,\\
\left\{{\cal N},\dot{x}^1\right\}&=&\lambda(\dot{x})\left\{{\cal N},-2 p_1-2\ell p_0 p_1\right\}=-\dot{x}^0\,,
\end{eqnarray}
and then that Finsler's $\dot{x}$ coordinate close a simple Lie algebra (while momenta obey to more complicated relations).
} we should remember this feature later, when we will discuss the line-element redefinition.

On the base of the just derived results we propose to identify the invariant  of our Lagrangian theory by subtracting the aforementioned boundary terms to (\ref{eq:dSFinslerLagrangian}). Indeed, using (\ref{pdeltaxpunto2}) and (\ref{puzzle1}) one gets:
\begin{equation}
\delta_{\cal N}{\cal L}-p_\mu \delta_{\cal N}\dot{x}^\mu=\delta_{\cal N}{\cal L}-\frac{\ell}{2}\dot{x}^1\left(p_1^2+2 p_0^2\right)=\delta_{\cal N}\left({\cal L}+m\left(2m\lambda(\dot{x})-F(\dot{x})\right)\right)=\delta_{\cal N}(2m^2\lambda(\dot{x}))\,.\label{puzzle2}
\end{equation}
Now, if we recall $\lambda(\dot{x})$ parameter definition (\ref{lambda}):
\begin{eqnarray}
\lambda(\dot{x})&=&\frac{1}{2}\frac{\sqrt{(\dot{x}^0)^2-(\dot{x}^1)^2}}{m}+
\frac{\ell}{2}\frac{\dot{x}^0(\dot{x}^1)^2}{(\dot{x}^0)^2-(\dot{x}^1)^2}\simeq
\frac{1}{2 m} \sqrt{(\dot{x}^0)^2 - \Bigg( 1 - 2\ell\frac{m \dot{x}^0}{\sqrt{(\dot{x}^0)^2-(\dot{x}^1)^2}}\Bigg) (\dot{x}^1)^2}=\nonumber\\
&=&\frac{\sqrt{\zeta_{\mu\nu}(\dot{x})\dot{x}^\mu\dot{x}^\nu}}{2m}\,,
\end{eqnarray}
we can finally identify the expression of our invariant Lagrangian (\ref{puzzle2}) as
\begin{equation}
\mathcal{L}_{inv}=m\sqrt{\zeta_{\mu\nu}(\dot{x})\dot{x}^\mu\dot{x}^\nu}\,.
\end{equation}
Here $\zeta_{\mu\nu}(\dot x)$ is the inverse metric of de Sitter momentum space defined in (\ref{eq:momSpacemetric}), written in terms of velocities by use of eqs. (\ref{eq:p0_complete}) and (\ref{eq:p1_complete}).

Notice that this Lagrangian cannot be directly related to a proper Finsler norm, since the associated metric $\zeta_{\mu\nu}(\dot x)$ does not satisfy the property in eq.(\ref{eq:propertiesofgF}).
Nevertheless, we can use this Lagrangian to derive the equations of motion for a particle. One gets again the geodesic equation (\ref{eq:geodesicEQ}), where now the Christoffel symbols are computed using $\zeta_{\mu\nu}(\dot x)$.
However, the difference between the equations of motion obtained using the two metrics $g_{\mu\nu}$ and $\zeta_{\mu\nu}$ can be reabsorbed by changing the normalization of the affine parameter $\tau$, which indeed was fixed by $g_{\mu\nu}\dot x^\mu \dot x^\nu=1$ in one case and $\zeta_{\mu\nu}\dot x^\mu \dot x^\nu=1$ in the other case.
This can be shown also by computing the worldlines: they are the same in the two cases  once one writes the coordinate velocity in terms of momenta $p$ instead of velocities $\dot x$. When the coordinate velocity is written in terms of $\dot x$ the two worldlines have a different form, since the definition of $\dot x\equiv\frac{d x}{d\tau}$ also depends on the normalization of the affine parameter.

Let us conclude this section with a brief summary about the different roles of the two metrics $g$ and $\zeta$.
It is known \cite{AmelinoCamelia:2011bm, kowalskiDSR2003,KowalskiGlikman:2003we, KowalskiGlikman:2004tz} that the sort of scenarios for  
DSR-deformed relativistic symmetries here considered
imply non-trivial properties of momentum space. In particular the symmetries encoded in the $\kappa$-Poincar\'e group have been described in terms of a curved momentum space with de Sitter metric, the metric we here
denoted by $\zeta^{\mu\nu}(p)$ (see end of section \ref{sec:kappaPoincare}). As mentioned at the end of subsection \ref{sec:FinslerMetric}, this metric allows to derive the dispersion relation of a particle whose symmetries are the ones of $\kappa$-Poincar\'e by computing the geodesic distance from the origin of momentum space to the point $p$, where $p$ is the momentum of the particle. It was shown in \cite{GiuliaFlaviokappaLocality, Trevisan} that one actually obtains a dispersion relation which is a function of the one we report in  eq. (\ref{eq:dSdispersionRelation}).
The momentum space metric allows to define an invariant momentum-space line element as:
\be
ds_p^2=\zeta^{\mu\nu}(p)dp_\mu dp_\nu,
\ee
and its inverse allows to define an invariant spacetime line element, which is invariant under the $\kappa$-Poincar\'e symmetries as shown earlier in this section\footnote{see also \cite{Loret:2014uia} for a more in depth discussion of the  of this line element}:
\be
ds^2=\zeta_{\mu\nu}(p)dx^\mu d x^\nu.
\label{gac2}
\ee

The definition of this momentum space metric is made easy by the fact that we are studying a case where spacetime is flat \footnote{Also the procedure described in this section to build an invariant line element relies on the flatness of spacetime.}.
When going to the more general case where curvature is present in both spacetime and momentum space (indicating that the local symmetry group of the geometry is a deformation of the Poincar\'e group), it is  not possible to separately describe spacetime and momentum space, and one has to rely on a geometrical structure that encodes at once the properties of the full phase space. In light of what we have found in this work, we conjecture that this structure should be provided by a Finsler metric.
Indeed, in this work we have shown that it is  possible to construct a Finsler metric compatible with the deformed relativistic symmetries of $\kappa$-Poincar\'e. While the metric $\zeta$ is a metric on momentum space, the Finsler metric $g$ is the proper generalization of the spacetime metric we are used to in the context of general relativity. The velocity dependence of $g$ encodes the non-trivial properties of spacetime induced by the deformed symmetry group, and allows to construct a geometry on the full phase space (it also naturally allows for the introduction of spacetime curvature).
The metric $g_{\mu\nu}$ 
allows to express the Casimir of the deformed symmetries in very simple manner by use of its inverse\footnote{Again, we write here explicitly the possible $x$ dependence of the metric to stress that it is defined on the full phase space.}:
\be
\mathcal C(p)=g^{\mu\nu}(x,p) p_\mu p_\nu.
\label{gac1}
\ee
It also allows to compute particles worldlines by use of the geodesic equation, despite the fact that the line element one could naively build with this metric is not invariant.

\section{Conservation laws in interactions}

Until now we have considered the Finsler geometry of a non-interacting single particle. The introduction of interactions goes beyond the scopes of this work, but we can still discuss what kind of composition laws for momenta are allowed within a given Finsler geometry. The issue is non-trivial, as the composition law of momenta has to be covariant under the deformed symmetries given by the Killing vectors.
This is already know in the framework of $\kappa$-Poincar\'e, where one has to introduce a deformed addition rule, $p\oplus q$, that has been shown to be related to the coproduct of translation generators.

Here we will investigate which deformations of the composition law are allowed within a given Finsler geometry associated to a deformed dispersion relation. In particular we will concentrate on the case studied here in section \ref{sec:FinslerBiscross}, with the MDR inspired by $\kappa$-Poincar\'e in the bicrossproduct basis.
We have seen that there exists a family of deformed boost that are compatible with the modified dispersion relation, and this family can be parameterized as in eq. (\ref{eq:NkappaCompatible}), which we rewrite here for convenience:
\be
\mathcal N_{\kappa-compatible}= p_1 x^0 + p_0 x^1+\ell\left(\alpha p_{0}p_{1}x^{0}+\gamma (p_{0}p_{1}x^{1}+ p_{1}^{2}x^{0})+(\alpha-1) p_{0}^{2}x^{1}-\frac{1}{2} p_{1}^{2}x^{1} \right)
\ee
Note that we have set $A=B=0$ in the boost representation, as the terms they multiply have no role in boosting a momentum.

We parameterize the most generic first-order deformation of the composition law as follows:
\bea
(p\oplus q)_{0}&=&p_{0}+q_{0}+\ell({\cal A} p_{0}q_{0}+{\cal B} p_{1}q_{1})\\
(p\oplus q)_{1}&=&p_{1}+q_{1}+\ell({\cal C} p_{1}q_{0}+{\cal D} p_{0}q_{1})
\eea
where the only conditions we asked for are that $p\oplus 0=0\oplus p=p$ and that the vector indices are coherent.

We look for the constraints on ${\cal A, B, C, D}$ such that the composition law is covariant under the action of the boost.

In order to ensure  relativistic compatibility between the boost and the composition law we ask that, if $(p\oplus q)_{\mu}=k_{\mu}$, then
\be
(p\oplus q)'_{\mu}\left\{\begin{array}{l}=k_{\mu}'\equiv k_{\mu} +\xi  \{{\cal N}_{\kappa-compatible}, k_{\mu}\}\\=(p'\oplus q')_{\mu}=\left[ ( p +\xi  \{{\cal N}_{\kappa-compatible}, p\})\oplus (q +\xi  \{{\cal N}_{\kappa-compatible}, q\})\right]_{\mu}\end{array}\right. \label{eq:SimpleCovariance}
\ee
where $\xi$ is the rapidity parameter.

The conditions we obtain are:
\bea
&& \gamma=0\\
&& {\cal A} = 0 \\
&& {\cal B} = 2 \alpha - 1\\
&& {\cal C} = \alpha - 1\\
&& {\cal D} = \alpha - 1
\eea

which means that we have a one-to-one correspondence between boosts and composition rules and the freedom in fixing the couple boost/composition rule, given a dispersion relation, is encoded in only one free parameter.

In the bicrossproduct basis of $\kappa$-Poincar\'e the composition rule reads:
\bea
(p\oplus q)_{0}&=&p_{0}+q_{0}\\
(p\oplus q)_{1}&=&p_{1}+(1-\ell p_{0})q_{1}
\eea
It has been shown \cite{GiuliaFlaviokappaLocality, AmelinoCamelia:2011yi} that this composition law is covariant under the action of the $\kappa$-Poincar\'e group in a peculiar way: given $(p\oplus q)_{\mu}=k_{\mu}$, then under an infinitesimal boost
\be
(p\oplus q)'_{\mu}\left\{\begin{array}{l}=k_{\mu}'\equiv k_{\mu} +\xi  \{{\cal N}_{\kappa}, k_{\mu}\}\\=(p'\oplus q')_{\mu}=\left[ ( p +\xi  \{{\cal N}_{\kappa}, p\})\oplus (q +\xi(1-\ell p_{0})  \{{\cal N}_{\kappa}, q\})\right]_{\mu}\end{array}\right.
\ee
The nontrivial feature is the deformation of the rapidity associated to the second momentum in the sum ($\xi\rightarrow \xi(1-\ell p_{0})$), with the deformation depending on the first momentum. It has been discussed in previous works \cite{AmelinoCamelia:2011yi, Amelino-Camelia:2013sba} how this deformation does not spoil the relativistic properties of the composition law.

The possibility of having this peculiar transformation law for particles entering into a vertex, such that the rapidity with which each particle is boosted depends on the momenta of the other particles in the vertex, allows for a widening of the possible composition rules/deformed boosts that are compatible with a given deformed dispersion relation.

To show this we generalize the covariance condition (\ref{eq:SimpleCovariance}) to:

\be
(p\oplus q)'_{\mu}\left\{\begin{array}{l}=k_{\mu}'\equiv k_{\mu} +\xi  \{{\cal N}_{\kappa-compatible}, k_{\mu}\}\\=(p'\oplus q')_{\mu}=\left[ ( p +\xi_{1}  \{{\cal N}_{\kappa-compatible}, p\})\oplus (q +\xi_{2}  \{{\cal N}_{\kappa-compatible}, q\})\right]_{\mu}\end{array}\right. \label{eq:MomentumDependentCovariance}
\ee
where $\xi_{1}=\xi(1+\ell( f_{11}q_{0}+f_{12}q_{1}))$ and $\xi_{2}=\xi(1+\ell (f_{21}p_{0}+f_{22}p_{1}))$, with $f_{ij}$ numerical coefficients.

The compatibility conditions we obtain are:
\bea
&& f_{12}=f_{22}=\gamma\\
&& {\cal A} = 0 \\
&& {\cal B} = 2 \alpha - 1 - f_{11} - f_{21} \\
&& {\cal C} = \alpha - 1 - f_{21} \\
&& {\cal D} = \alpha - 1 - f_{11}
\eea

So we have a total of four free parameters.
This means that if we fix completely the form of the boost (three parameters), we still have one free parameter left, which represents a freedom in the choice of the composition rule.
Note that it is still impossible to have a standard composition rule $p \oplus q=p + q$.

\section{Conclusions and outlook}

In summary we have shown that there is a well defined relationship between modified relativistic symmetries and Finsler geometries. Starting from a dispersion relation which is inspired by the Casimir of $\kappa$-Poincar\'{e} in a given basis, we calculated the correspondent Finsler geometry and showed that the latter provides the same worldlines as in $\kappa$-Poincar\'{e}. Then, we considered the conserved charges associated to isometries in the Finsler geometry and showed that there exists one choice of the free parameters and functions in the Killing--Finsler symmetries that reproduces the $\kappa$-Poincar\'e conserved charges.
Furthermore, we showed that the additional freedom provided by the free functions appearing in the Finsler conserved charges mirrors the freedom that one has to redefine the $\kappa$-Poincar\'e symmetry generators without spoiling the invariance of the Casimir. While these results were initially proven in a special basis of $\kappa$-Poincar\'e (the so called bicrossproduct basis), we have seen that
upon a (nonlinear) redefinition of the translation generators in the $\kappa$-Poincar\'{e} algebra, and hence upon the correspondent change in the Casimir, the new Finsler geometry still allows to reproduce the features of $\kappa$-Poincar\'{e} in this new basis by recovering, in particular, the correct form of associated conserved charges. We have also discussed how to redefine, by a boundary term that leaves the physical quantities unchanged, the Lagrangian so to have it conserved under boosts. The geodesics can in this case be seen as those of an auxiliary metric $\zeta(\dot{x})$, and are the same curves as those derived from the Finsler geometry $g(\dot{x})$ upon a suitable change of the normalization of the affine parameter. Finally, we have elaborated on the possible generalization of the framework to particles' interactions.

We think that the above mentioned results are clearly suggestive of a deep link between deformed relativistic groups and Finsler geometries, i.e. geometrical characterizations of the phase space structure.
One might wonder how this could be the case. While the present investigation falls short of enlightening the physical reasons for this link, it is perhaps possible to speculate how this might arise.
A special relativistic structure is rooted in very basic assumptions about the structure of space and time (see e.g.~\cite{Liberati:2013xla} for a review of the axiomatic derivation of Special Relativity): pre-causality (invariance of time ordering of co-local events in any reference frame), the relativity principle (equivalence of inertial reference frames), isotropy of space and homogeneity of space-time. Searching for possible UV departures from this scheme without violation of the relativity principle leaves substantially only the option to relax isotropy or homogeneity. Isotropy-breaking relativity groups have been already considered in the literature \cite{Cohen:2006ky,Gibbons:2007iu,Sonego:2008iu} and proven to be described by Finslerian line elements (which are invariant under symmetry groups with less generators than in special relativity, at least in more than 1+1 dimensions). Homogeneity departures are far less explored. However, as noticed in \cite{DiCasola:2014xya}, it is easy to see that relaxing homogeneity of space-time is tantamount to renounce to an operative meaning of coordinates (in the sense that differences of spatial and time coordinates are no more interpretable respectively as lengths and durations) typical of a special relativistic framework. We conjecture that this breakdown of the operative meaning of coordinates is at the root of necessity to describe physical phenomena in a full phase space given that in this case velocities cannot be trivially derived as a limiting procedure of the ratio of $\Delta x/\Delta t$. If this conjecture will be proven correct the implications would be striking as they would suggest that between our IR reality and the UV, full quantum gravity regime (where a continuous spacetime geometry can be completely absent), there would generically lie a mesoscopic regime where a full fledged phase space-based description of physical phenomena is needed.

In this sense the work here performed is susceptible of interesting developments as it would naturally allow for generalizations to curved spacetimes of previous DSR-scenarios investigations (in alternative or together with curved momentum space structures) which might be applied to long standing problems in theoretical physics (as e.g.~black hole physics). Also it would be interesting to study how in quantum gravity approaches the spacetime metric can acquire a dependence on the typical velocities or momenta at short scales so to lead to the Finslerian structures here discussed. We limit ourselves in noticing here that renormalization group approaches applied to gravity \cite{Reuter:2012id} seems to naturally lean towards these scenarios. We hope to come back on this and other issues in future investigations.

\begin{acknowledgments}
This work was supported in part by a grant from the John Templeton Foundation.
\end{acknowledgments}

\appendix

\section{Derivation of Killing equation in Finsler geometry}
\label{sec:DerivationOfKilling}
It is useful, in order to better understand the discussion on symmetries, to derive explicitly the Killing equation in Finsler geometry.\\
In Finsler spacetime we can express the variation of the coordinates $x^\alpha$ along a vector field $\xi^\alpha$ as:
\begin{equation}
(x^\prime)^\alpha=x^\alpha+\xi^\alpha \delta \lambda\,,\label{varx}
\end{equation}
where $\lambda$ is the infinitesimal variation parameter. This variation of $x^\alpha$ reflects on $\dot{x}^\alpha$ in the following way
\begin{equation}
(\dot{x}^\alpha)^\prime=\dot{x}^\alpha+\frac{\partial \xi^\alpha}{\partial x^\beta}\frac{d x^\beta}{d\tau} \delta \lambda = \dot{x}^\alpha+\frac{\partial \xi^\alpha}{\partial x^\beta}\dot{x}^\beta \delta \lambda\,.
\end{equation}
The general variation of a vector field $X^\alpha(x,\dot{x})$ will then be
\begin{equation}
\delta X^\alpha=\frac{\partial X^\alpha}{\partial x^\beta}\xi^\beta\delta\lambda + \frac{\partial X^\alpha}{\partial \dot{x}^\gamma}\frac{\partial \xi^\gamma}{\partial x^\beta}\dot{x}^\beta\delta\lambda\,.\label{varvec}
\end{equation}
As in general relativity, in Finsler geometry we can obtain the Killing equation by imposing the line-element invariance with respect to the variation along a vector field $\xi^\alpha$:
\begin{equation}
\delta(ds^2)=\delta(g_{\mu\nu}dx^\mu dx^\nu)=\delta (g_{\mu\nu})dx^\mu dx^\nu + g_{\mu\nu}\left(\delta(dx^\mu)dx^\nu + dx^\mu \delta(dx^\nu)\right)=0\,.\label{linelinv}
\end{equation}
From (\ref{varvec}) we know that:
\begin{equation}
\delta(g_{\mu\nu})=\partial_\alpha g_{\mu\nu}\xi^\alpha\delta\lambda + \frac{\partial g_{\mu\nu}}{\partial\dot{x}^\beta}\partial_\alpha\xi^\beta \dot{x}^\alpha\delta\lambda\,,
\end{equation}
while from (\ref{varx}) we can obtain:
\begin{eqnarray}
\delta(dx^\alpha)=d(\delta x^\alpha)=d(\xi^\alpha\delta\lambda)=\partial_\beta \xi^\alpha dx^\beta \delta\lambda\,.
\end{eqnarray}
Therefore equation (\ref{linelinv}) can be expressed as:
\begin{equation}
\delta(ds^2)=\left( \partial_\alpha g_{\mu\nu}\xi^\alpha + \frac{\partial g_{\mu\nu}}{\partial\dot{x}^\beta}\partial_\alpha\xi^\beta \dot{x}^\alpha \right)dx^\mu dx^\nu + g_{\mu\nu}\left(\partial_\beta \xi^\mu dx^\beta dx^\nu + dx^\mu \partial_\beta \xi^\nu dx^\beta\right)=0\,.
\end{equation}
In the end we find the generalized Killing equation we used in (\ref{eq:Killing1}):
\begin{equation}
 \partial_\alpha g_{\mu\nu}\xi^\alpha + g_{\alpha\nu}\partial_\mu \xi^\alpha +  g_{\mu\alpha}\partial_\nu \xi^\alpha + \frac{\partial g_{\mu\nu}}{\partial\dot{x}^\beta}\partial_\alpha\xi^\beta \dot{x}^\alpha  =0\,.\label{KillingEq}
\end{equation}


\begin{thebibliography}{99}

\bibitem{GACijmpd2002}
G.~Amelino-Camelia,
  Int.\ J.\ Mod.\ Phys.\ D {\bf 11} (2002) 35
  [gr-qc/0012051].

\bibitem{smolinDSR2003}
J.~Magueijo and L.~Smolin,
  Phys.\ Rev.\ D {\bf 67} (2003) 044017
  [gr-qc/0207085].

\bibitem{kowalskiDSR2003}
 J.~Kowalski-Glikman,
  Phys.\ Lett.\ B {\bf 547} (2002) 291
  [hep-th/0207279].


\bibitem{gacSYMMETRYreview}
G.~Amelino-Camelia,
  Symmetry {\bf 2} (2010) 230
  [arXiv:1003.3942 [gr-qc]].

\bibitem{MDRFinsler}
F.~Girelli, S.~Liberati and L.~Sindoni,
  Phys.\ Rev.\ D {\bf 75} (2007) 064015
  [gr-qc/0611024].


\bibitem{GiuliaFlaviokappaLocality}
 G.~Gubitosi and F.~Mercati,
  Class.\ Quant.\ Grav.\  {\bf 30} (2013) 145002
  [arXiv:1106.5710 [gr-qc]].

\bibitem{Trevisan}
G.~Amelino-Camelia, M.~Arzano, J.~Kowalski-Glikman, G.~Rosati and G.~Trevisan,
  Class.\ Quant.\ Grav.\  {\bf 29} (2012) 075007
  [arXiv:1107.1724 [hep-th]].



 \bibitem{Lukierski:1992dt}
  J.~Lukierski, A.~Nowicki and H.~Ruegg,
  Phys.\ Lett.\ B {\bf 293} (1992) 344.

\bibitem{Lukierski:1993wxa}
  J.~Lukierski and H.~Ruegg,
  Phys.\ Lett.\ B {\bf 329} (1994) 189
  [hep-th/9310117].

  \bibitem{Lukierski:1991pn}
  J.~Lukierski, H.~Ruegg, A.~Nowicki and V.~N.~Tolstoi,
  Phys.\ Lett.\ B {\bf 264} (1991) 331.

\bibitem{Freidel:2005me}
  L.~Freidel and E.~R.~Livine,
  Phys.\ Rev.\ Lett.\  {\bf 96} (2006) 221301
  [hep-th/0512113].

\bibitem{MajidBicross}
  S.~Majid and H.~Ruegg,
  Phys.\ Lett.\ B {\bf 334} (1994) 348
  [hep-th/9405107].

\bibitem{lateshift}
 G.~Amelino-Camelia, L.~Barcaroli, G.~Gubitosi and N.~Loret,
  Class.\ Quant.\ Grav.\  {\bf 30} (2013) 235002
  [arXiv:1305.5062 [gr-qc]].


  \bibitem{Lukierski:1993wx}
  J.~Lukierski, H.~Ruegg and W.~J.~Zakrzewski,
  Annals Phys.\  {\bf 243} (1995) 90
  [hep-th/9312153].

 \bibitem{Rosati:2012fb}   G.~Rosati, N.~Loret and G.~Amelino-Camelia,   
  J.\ Phys.\ Conf.\ Ser.\  {\bf 343} (2012) 012105
  [arXiv:1203.4677 [hep-th]].

\bibitem{AmelinoCamelia:2010qv}   G.~Amelino-Camelia, M.~Matassa, F.~Mercati and G.~Rosati,   
  Phys.\ Rev.\ Lett.\  {\bf 106} (2011) 071301
  [arXiv:1006.2126 [gr-qc]].

\bibitem{kbob}
G.~Amelino-Camelia, N.~Loret and G.~Rosati,
  Phys.\ Lett.\ B {\bf 700} (2011) 150
  [arXiv:1102.4637 [hep-th]].


\bibitem{KowalskiGlikman:2003we}
  J.~Kowalski-Glikman and S.~Nowak,
  Class.\ Quant.\ Grav.\  {\bf 20} (2003) 4799
  [hep-th/0304101].

  \bibitem{KowalskiGlikman:2004tz}
  J.~Kowalski-Glikman and S.~Nowak,
  hep-th/0411154.

\bibitem{AmelinoCamelia:2011bm}
  G.~Amelino-Camelia, L.~Freidel, J.~Kowalski-Glikman and L.~Smolin,
  Phys.\ Rev.\ D {\bf 84} (2011) 084010
  [arXiv:1101.0931 [hep-th]].


  \bibitem{HRund}
H. Rund,
 {\it The differential geometry of Finsler spaces.}
 Springer, Berlin (1959).

\bibitem{MignemiFinsler}
S.~Mignemi,
  Phys.\ Rev.\ D {\bf 76} (2007) 047702
  [arXiv:0704.1728 [gr-qc]].

\bibitem{Loret:2014uia}
  N.~Loret,
  arXiv:1404.5093 [hep-th].


\bibitem{AmelinoCamelia:2011yi}
  G.~Amelino-Camelia,
  Phys.\ Rev.\ D {\bf 85} (2012) 084034
  [arXiv:1110.5081 [hep-th]].

\bibitem{Amelino-Camelia:2013sba}
  G.~Amelino-Camelia, G.~Gubitosi and G.~Palmisano,
  arXiv:1307.7988.

\bibitem{Liberati:2013xla}
  S.~Liberati,
  Class.\ Quant.\ Grav.\  {\bf 30}, 133001 (2013)
  [arXiv:1304.5795 [gr-qc]].

\bibitem{Cohen:2006ky}
  A.~G.~Cohen and S.~L.~Glashow,
  Phys.\ Rev.\ Lett.\  {\bf 97}, 021601 (2006)
  [hep-ph/0601236].

\bibitem{Gibbons:2007iu}
  G.~W.~Gibbons, J.~Gomis and C.~N.~Pope,
  Phys.\ Rev.\ D {\bf 76}, 081701 (2007)
  [arXiv:0707.2174 [hep-th]].

\bibitem{Sonego:2008iu}
  S.~Sonego and M.~Pin,
  J.\ Math.\ Phys.\  {\bf 50}, 042902 (2009)
  [arXiv:0812.1294 [gr-qc]].

\bibitem{DiCasola:2014xya}
  E.~Di Casola, S.~Liberati and S.~Sonego,
  arXiv:1405.5085 [gr-qc].

\bibitem{Reuter:2012id}
  M.~Reuter and F.~Saueressig,
  New J.\ Phys.\  {\bf 14}, 055022 (2012)
  [arXiv:1202.2274 [hep-th]].



\end{thebibliography}
\end{document}